# Placement of Biological Membrane Patches in a Nanofluidic Gap with Control over Position and Orientation

*Francesca Ruggeri[1,a], Christian Schwemmer[1,a], Mirko Stauffer[2], Philippe M. Nicollier[1], Jacqueline Figueiredo da Silva[1], Patrick D. Bosshart[2], Kirstin Kochems[3], Dimitrios Fotiadis[2], Armin Knoll[1], Heiko Wolf[1] \**

[1]IBM Research – Zurich, Säumerstrasse 4, 8803 Rüschlikon, Switzerland
[2]Institute of Biochemistry and Molecular Medicine, University of Bern, 3012 Bern, Switzerland
[3]Experimental Physics and Center for Biophysics, Saarland University, 66123 Saarbrücken, Germany
[a] contributed equally
\* corresponding author: hwo@zurich.ibm.com

**Abstract**

Purple membranes from the archaeon *Halobacterium salinarum* consist of two-dimensional crystals of the light-driven proton pump bacteriorhodopsin, which convert photons into a proton gradient across the cell membrane. This functional feature and the structural rigidity make them appealing candidates for integration into biomimetic devices. To this end, and in order to carry out their function, purple membranes must be positioned in the correct orientation at the position of interest. Precise placement and control over the orientation of nanoscale objects still constitutes a formidable challenge. Here we show that isolated purple membrane patches can be transported and positioned at predefined locations in nanofluidic confinement, with control over their orientation at the target sites. The transport is achieved through a rocking Brownian motor scheme, while the controlled deposition of the membranes is realized by engineering the surface potential of a fluid-filled nanofluidic slit. This controlled manipulation of purple membrane patches outlines a new pathway towards the integration of biological or other delicate supramolecular structures into top-down-fabricated patterns, for the assembly of nanoscale hybrid devices that serve as a light-driven source of (chemical) energy.

**Introduction**

A component that converts radiant energy into chemical energy is a desirable building block for nanoscale devices.[1] Such devices might be applied as artificial chemical nanoreactors where they can drive and fuel chemical reaction cycles or chemical reaction networks.[2] Light-driven chemical signaling is also conceivable.[3] Certain transmembrane proteins can supply such functionality. The membrane protein bacteriorhodopsin (bR) is an example of a light-driven proton pump that is found in very high concentrations in biological membrane patches, called purple membranes (PMs), of the archeon *Halobacterium salinarum* (previously known as *Halobacterium halobium*).[4]

PMs consist of about three quarters of bR and one quarter of lipids that occupy the interstitial spaces between the proteins.[4,5] bR forms trimers, which are densely packed in a hexagonal lattice, as first discovered by X-ray diffraction and electron microscopy,[6,7] and later also observed by atomic force microscopy (AFM).[8] Isolated PM patches typically have dimensions of 500 nm (diameter) but size can vary depending on preparations. The dense packing of bR renders purified PM patches fairly stiff[9] and shapes them into flat platelets. In a typical photocycle, bR undergoes several spectroscopically detectable conformational changes during which a proton is pumped from the inside to the outside of the cell.[5] The established proton gradient is used by the bacterium to fuel the synthesis of adenosine triphosphate (ATP). Thus, the bR proton pumping activity is inherently directional. bR consists of seven transmembrane α-helices with N- and C-termini of the protein at the extracellular and cytoplasmic side, respectively. The structural and physicochemical composition of PMs (*e.g.*, amino acid side chains in the loops connecting the transmembrane α-helices and the charge of lipid head groups) predicts an overall negative charge with a much stronger negative potential on the C-terminal intracellular side.[10] However, experiments reveal a significantly smaller dipole than theoretically predicted.[10] This is supposed to be caused by bivalent cations that are tightly bound to some of the anionic sites in the membranes, screening part of the charged groups.[11] Even though results vary, a majority of experiments still finds more negative surface charge on the cytoplasmic side where the C-terminus of the bR protein is located.[10] Overall, the PM patches are negatively charged and stable in ultrapure water (Millipore) at pH 6 – 7, without additional buffering agents.

Given the inherent asymmetry in their functionality, an integration of the PMs in any kind of device necessitates control over their orientation. Surface adsorption of PMs with a preferential orientation has been previously demonstrated by engineering the substrate surface charge[12,13] or by functionalizing one side of the membranes with a specific linker.[14] However, control over the specific position of adsorption, at the level of the individual membrane patch, was not possible using those methods. The adsorption of PMs on $Si/SiO_2$[15] or $SiN_x$[16] nanopores yields both orientations.

Today, it is still a challenge to manipulate and integrate delicate biological building blocks into artificial devices without damage. To address this issue, a range of trapping methods have been developed over the years to control and study biological objects in a state comparable to their free state (*i.e.,* without surface tethering). The most common methods are based on polarizability-dependent external fields (*e.g.,* optical tweezers) that suffer from an unfavorable scaling of trap depth with the volume of the particle, with a limited range of applicability for small biomolecules. It has been shown[17] that high optical or electrical fields may even cause a disruption of the structural integrity of the object of interest (protein unfolding).

On the contrary, the Electrostatic Fluidic Trap principle, first demonstrated by Krishnan and collaborators,[18,19] has recently been applied to biomolecules with a charge magnitude ranging from 1 to $100e$ (where $e$ is the charge of one electron), and as small as a single fluorophore.[20] The typical Electrostatic Trap device comprises a nanofluidic slit of 100-200 nm gap height ($2h$), in which one side is made of glass, while the other consists of an oxide and is patterned with circular indentations of a typical depth, $d$, and diameter, $D$ (Figure 1a). The glass and oxide surfaces acquire a net charge when in contact with an aqueous solution: in order to maintain electroneutrality in the system the surface charge of the slit walls is balanced by an equal number of counterions of opposite charge, partially adsorbed to the surface in the Stern layer, and partially forming the diffuse electrical double-layer.[21] The distribution of counterions of valence $z$ at any coordinate $z_p$ in the vertical plane is governed by the Boltzmann formula,

$$n(z_p) = n_0 \exp[-ze\psi(z_p)/k_BT] \qquad \text{(Eqn. 1)},$$

where $\psi$ is the electrostatic potential and $n_0$ is the number density of molecules at the surface, $k_B$ is Boltzmann's constant and $T$ the temperature of the system. In the case of flat surfaces, the resulting 1D electrostatic potential, obtained by solving the Poisson-Boltzmann equation for $\psi(z_p)$, can be approximated by the linearized expression

$$\psi(z_p) = \psi_s \exp(-\kappa z_p) \qquad \text{(Eqn. 2)}.$$

Here $\kappa^{-1}$ is the Debye length, typically 10-50 nm, that is proportional to $1/\sqrt{c}$, where $c$ is the salt concentration in the solution. $\psi_s$ is an effective surface potential, related to the surface charge via the pH of the aqueous solution, as modeled by Behrens and Grier.[22] The potential at the nanofluidic slit midplane $\psi_m$, resulting from the superposition of the electrostatic potential arising from the two charged surfaces in the slit of gap height $2h$, is $\psi_m = 2\psi_s \exp(-\kappa h)$. If we create a circular indentation of a depth $d > \kappa^{-1}$ in one of the surfaces of the slit, the local potential minimum approaches zero. This perturbation of the local electrostatic landscape creates a potential well of depth $\psi_m$ for a point-like charged particle sampling the slit. The total electrostatic free energy $W$ for said point-object of charge $q$ is $q\psi_m$.[19] In general, stable trapping (for a time $t > 10$ ms) of molecules is achieved for $W \geq 3\ k_BT$.[20] In previous work, we measured trapping energies in the range of 3 to 7 $k_BT$, depending on the gap distance between two negatively charged glass/glass or glass/polymer surfaces, using 60 nm Au nanoparticles as test objects.[23] As highlighted by Eqn. 2, electrostatic trapping in a nanofluidic slit provides higher confinement at low salt concentration, $c < 1$mM, thereby maximizing $\kappa^{-1}$ (typically between 10 and 30 nm in our experiments) and thus $\psi_m$.

The nanofluidic apparatus is also capable of transporting objects to the desired trapping sites by so-called "Rocking Brownian Ratchets". These entail a combination of asymmetric electrostatic potentials and a net-zero oscillating force, externally applied, which create a flow-free transport of nano-objects in the slit.[24]

Here, we illustrate how our experimental method, generating electrostatic landscapes, provides the unprecedented ability to transport, trap and deposit PMs with control over both location and orientation, thus providing a path towards hybrid devices with precisely integrated biological components adding crucial functionality.

**Results and Discussion**

**Membrane trapping.** We used the nanofluidic confinement apparatus described earlier[24,25] for our experiments. In brief, the setup consists of an interferometric scattering microscopy (ISCAT) configuration integrated with a movable nanofluidic slit, controlled by a piezo-electric stage. Nanometer-precision in the control of the gap allowed us to finely tune the electrostatic potential (see Eqn. 2). The top cover of the nanofluidic slit is a cover glass containing a pillar with an area of 80 μm by 80 μm and a height of 30 μm, surrounded by 4 electrodes in a cross-shaped configuration. The lower surface of the nanofluidic slit is a substrate patterned with circular traps, fabricated using thermal scanning probe lithography (t-SPL).[26,27] In t-SPL, a heat-sensitive resist (polyphthalaldehyde, PPA) is selectively removed using a heated AFM tip, which also allows us to pattern complex 3D shapes on the nanoscale.[28,29] Patterns in the resist were transferred into a subjacent $SiO_2$ layer by reactive ion etching (RIE) (for details see Methods). We prepared arrays of 9 by 9 (or 6 by 7) circular holes of 1.2 – 1.5 μm diameter, $D$, and a depth, $d$, of 15-30 nm (Figure S1). In the array, the distance between the centers of the holes was 2 μm. Several arrays of traps were patterned with a distance of 50 μm to 100 μm from each other (Figure S2). This gave us the possibility to perform several runs of membrane deposition during one experiment by simply moving to another trap array.

Membranes were dispersed in ultrapure water at a concentration of 0.1 – 0.2 mg ml$^{-1}$ without any further addition of salts or buffer. For the trapping experiments, 25 μl of membrane solution were placed on the patterned substrate already positioned in the nanofluidic setup. Then, the cover slide containing the glass pillar and the electrodes was lowered into contact with the drop, creating a slit of controlled gap height.

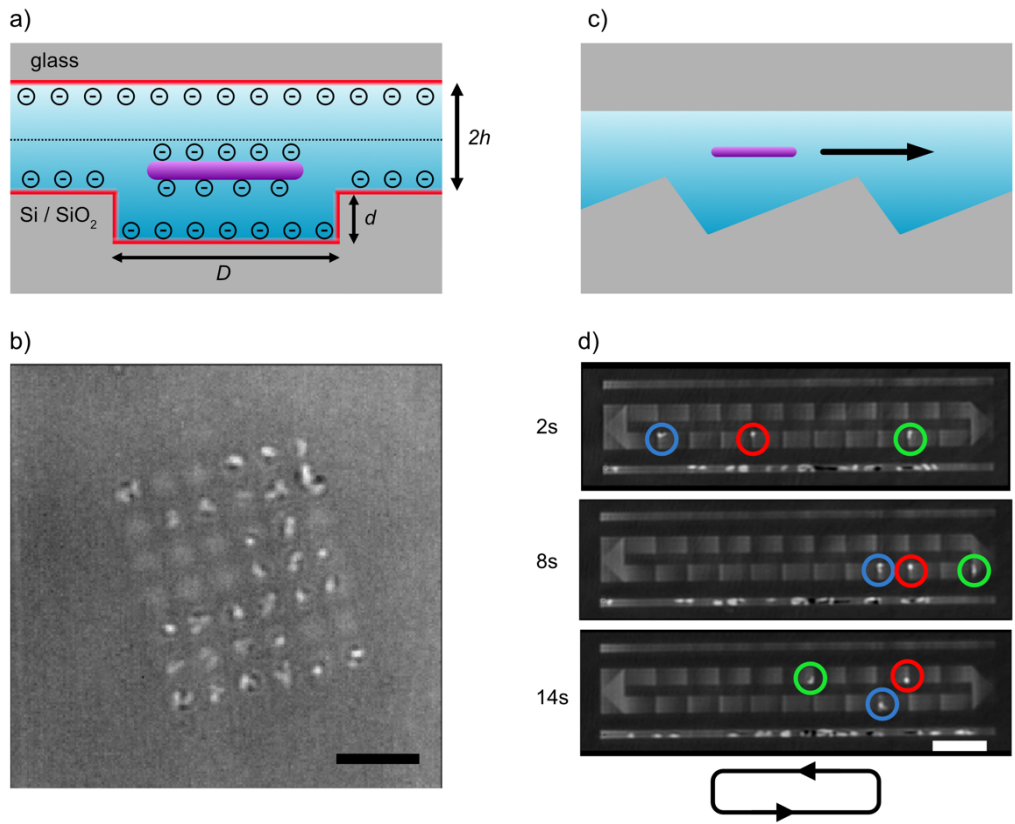

**Figure 1. Trapping and transport of PM patches in nanofluidic confinement.** (a) Scheme of a negatively charged PM trapped in nanofluidic confinement. (b) ISCAT image (contrast-enhanced, see Methods) of an array of traps with captured membranes. The membranes are clearly visible as bright spots with a darker edge. (c) Scheme of a PM being transported in a ratchet. (d) Screenshots from a movie (Movie S2) at different time steps, demonstrating the transport of membranes. The arrows indicate the direction of movement. The colors mark the same membrane in each of the frames. Scale bars are 5 μm.

Despite of their thickness of only about 5 - 6 nm, the PM patches produced a contrast in ISCAT which allowed us to image the PMs in the nanofluidic gap and track their position. As expected given their smaller scattering cross-section, the contrast of the ISCAT images was less pronounced than that observed for 60 nm gold nanoparticles (NPs).[30] A frame directly taken from a short movie (Movie S1) shows the array of traps with some traps having captured a PM patch, however, with low contrast (Figure S3a). Subtraction of the static background (for the detailed procedure see Methods) enhanced the contrast of PMs against the trap floor by a factor of about three (from 14% to 45%) and yielded an image with the membranes clearly discernable (Figure 1b), thus, demonstrating that label-free optical imaging of individual PM patches is possible with the ISCAT setup.

When sampling an unpatterned area of the bottom substrate, or when the gap height was large (hence, $\psi$ small) PMs were floating freely, undergoing Brownian motion. The membranes did not show unspecific adsorption to the surfaces. When the nanofluidic gap was reduced to about 200 nm, the membranes were repelled by both like-charged surfaces and preferentially stayed in the circular holes (*i.e.,* they were "trapped"), as schematically shown in Figure 1a. We point out that, although their location was confined, the membranes were still thermally fluctuating and were not adsorbed to the substrate (Movie S1). Upon increasing the gap height, the potential barriers decreased ($W$ < 5 kT), allowing the PMs to jump between traps or even escape the array.

Typically, we transported membranes towards the patterned area and into the potential wells using an electroosmotic flow produced by an AC electric field with a DC bias (0.5 V AC, 0.5 V DC bias) applied to the electrodes. By adjusting the gap height with the movable pillar and with the aid of the DC bias, it was also possible to direct a membrane from one neighboring trap to another within a given array. The cross-shaped electrode configuration on the pillar was used to define the direction of motion along the x or y direction of the array.

**Membrane transport.** The fact that the membranes can be both trapped and directed by an electroosmotic flow indicates that they might be prone to directional transport in an asymmetric ratchet potential. Thus, we tested their behavior in a rocking Brownian motor.

We prepared ratchet patterns in PPA using t-SPL (scheme Figure 1c). For the experiments with PMs, we used the ratchet patterns in PPA without further transfer into the oxide layer. Note that ratchets patterned in PPA can also be transferred into the substrate material when an integration with other patterns (e.g. traps) is required.[31] This transfer will not change the function of the device if the substrate attains a similar charge as the polymer. The experimental setup and instrument were the same as described for Au NPs in an earlier publication,[24] only the lateral dimensions of the ratchet were larger to accommodate the larger diameter of the PM patches. The ratchet period was 3 μm, with a width of 1.5 μm. The maximum depth of the ratchet teeth was 20 nm (Figure S4). The rocking potential applied by the pillar electrodes had peak voltages of 1.0 - 1.7 V and frequencies between 0.5 and 2 Hz. In Figure 1d, snapshots that were taken at intervals of 6 s from a movie (Movie S2) are presented. It can be observed how the membranes in the ratchet move around the track formed by the ratchet teeth. The membranes in the ratchet have a drift velocity of 8.6 ± 3.1 μm s$^{-1}$ (forward tilt, 1 V, 1 Hz). Importantly, the results show that a directed transport of the biological membranes can be achieved without a net fluid flow.

We envision this capability as a powerful tool to transport individual membranes along predefined paths and to specific trapping sites, to generate complex networks in future experimental implementations.

**Membrane deposition.** PMs that are confined in the circular holes of the trap arrays can be permanently deposited by decreasing the nanofluidic gap to near contact ($2h \sim \kappa^{-1}$). In this configuration, the membranes approach the trap surface enough for Van der Waals forces to overcome electrostatic repulsion and hydration layer forces.[23] We note that after this 'move into contact' the membranes are permanently adsorbed to the surface, and do not re-disperse once the gap is opened again or the sample is dried. It is important to point out that the membranes in the traps did not experience any simultaneous mechanical contact to both the pillar and the trap bottom during deposition, as the geometrical depth of the indentations was always 10 – 25 nm larger than the thickness of the biological membranes (5-6 nm). The possibility of immobilizing the membranes enabled us to perform several deposition cycles on a given trap array by alternating between opening (replenishing the membrane concentration over the observed area) and reducing the nanofluidic gap. The analysis of deposited membranes was performed by AFM after removing the substrates from the confinement apparatus, removing excess liquid, and drying (Figure S5, S6a). In some cases, scanning electron microscopy (SEM) was used additionally for the analysis of deposited membranes (Figure S6b, c).

**Membrane orientation.** To control the orientation of the membranes prior to their deposition on the substrate, we engineered the surface potential at the top and bottom of the trapping sites independently, thus energetically favoring one or the other membrane orientation. To this end, we prepared two kinds of trapping substrates, one with all interfaces made of $SiO_2$ (Type A) (Figure 2a) and a second one with side walls made of $SiO_2$ but a bottom surface of $Al_2O_3$ (Type B) (Figure 2b).

In addition to unmodified 'wild type' PMs, we applied PMs with a deca histidine ($His_{10}$)-tag attached to the bR protein to determine and control membrane orientation. Previous studies of PM orientation relied on membrane features visible in high-resolution electron microscopy,[12] high-resolution AFM,[32] or antibodies that are specific to one side of the membranes.[33] Our $His_{10}$-tagged PMs come in two varieties with the His-tag attached either at the bR C-terminus (cytoplasmic side, C-His) or the bR N-terminus (extracellular side, N-His). The His-tag modification was produced by genetic engineering of a *H. salinarum* strain as described previously[34,35] and in the Methods section.

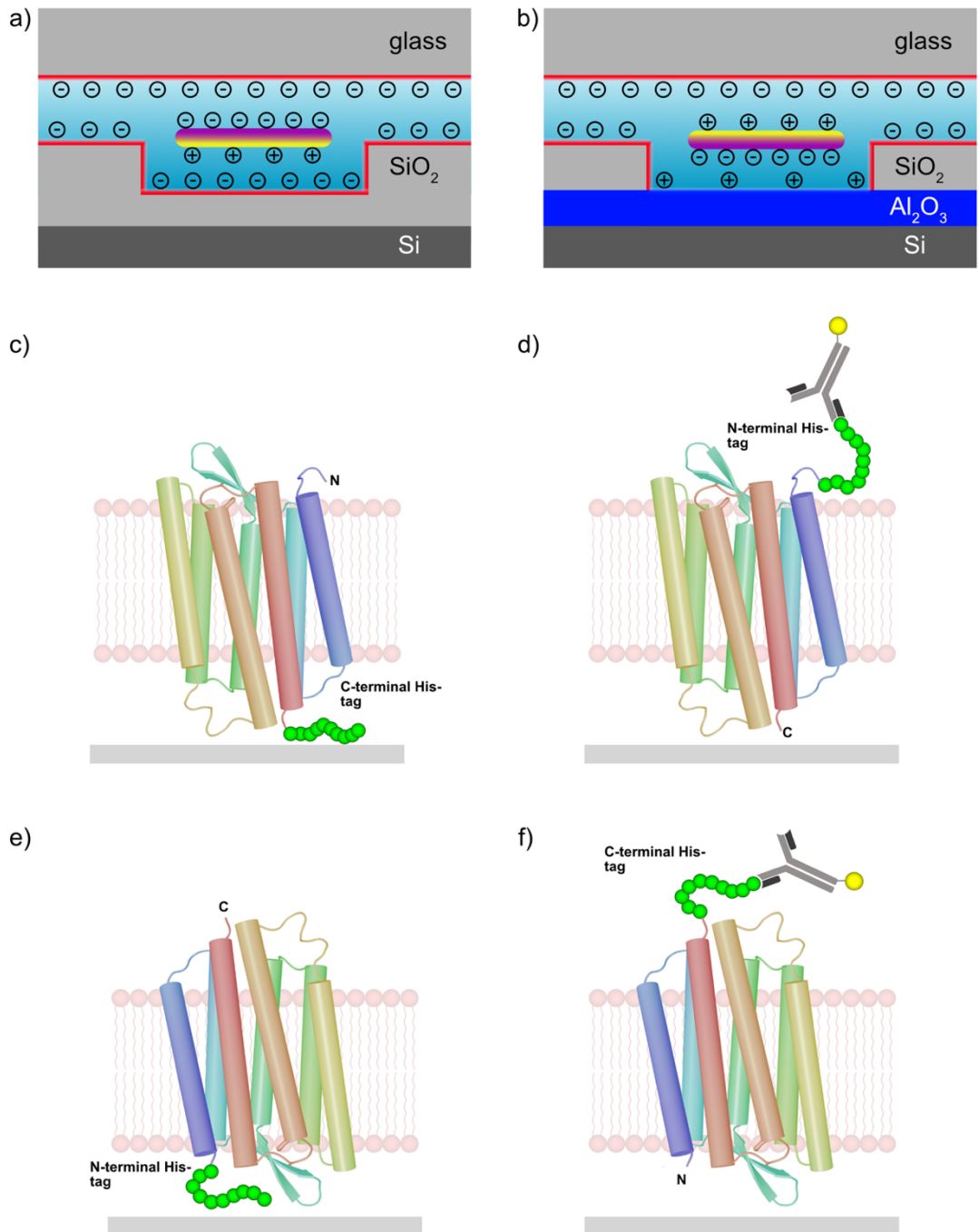

**Figure 2. Orientation of His-tag-modified PM patches in nanofluidic confinement and after deposition.** (a), (b) Scheme of an asymmetrically charged PM in nanofluidic confinement with its orientation dependent on the surface charge of the trap. (c)-(f) Scheme of bR in the membrane with a His-tag attached on C-terminus or N-terminus of bR, and an antibody attached depending on accessibility of the His-tag. Transmembrane α-helices of bR are displayed as cylinders and colored based on a rainbow coloring scheme. Protein termini are labeled with "N" (extracellular, N-terminus) and "C" (cytoplasmic, C-terminus), respectively. Membrane and lipids are shown in a faded red tone. The surface where bR is adsorbed is shown in light grey.

The His-tag served two purposes: (1) as an easy marker for the orientation detection of the PM patches and (2) to create an increased charge asymmetry between the different sides of the membrane patches.

For marking the His-tag, we used fluorescently labelled antibodies specific to the His-tag, which were applied after the deposition of the membranes. Only when PM patches were presenting their His-tag on top, the His-tag could be recognized by the antibody and the membranes showed fluorescence (schemes in Figure 2c-f). PM patches deposited with the His-tag towards the substrate surface were not accessible to the antibody and thus remained dark. A comparison of fluorescence images and AFM (or SEM) data of the same area revealed the ratio of orientation of the membranes (overlay of fluorescence image and SEM in Figure 3d).

For the charge distribution of the His-tagged PMs, the contribution of the His-tag itself to the surface charge of the membranes had to be considered. The pH of the unbuffered solutions of both C-terminal His-tagged and N-terminal His-tagged membranes was measured to be 6.3. The imidazole side chain of the histidine has a $pK_a$ of 6 [36] which means that in unbuffered solutions a significant number of His-tags carry a positive charge and that the His-tag contributes to an asymmetric charge distribution on the membranes.

In the described experiments, the addition of buffer was generally avoided to maintain a high Debye length and thereby a strong trapping potential (Eqn. 2). The use of 1 mM TRIS-HCl buffer at pH 8 decreases the Debye length to approximately 10 nm. Accordingly, stable trapping and membrane deposition is achievable at smaller gap distances.

After each deposition experiment, an antibody assay was performed. Control experiments with wild type PM patches (on Type B traps) showed no fluorescence from antibody binding, thus excluding any unspecific binding of the antibodies on the PM surfaces.

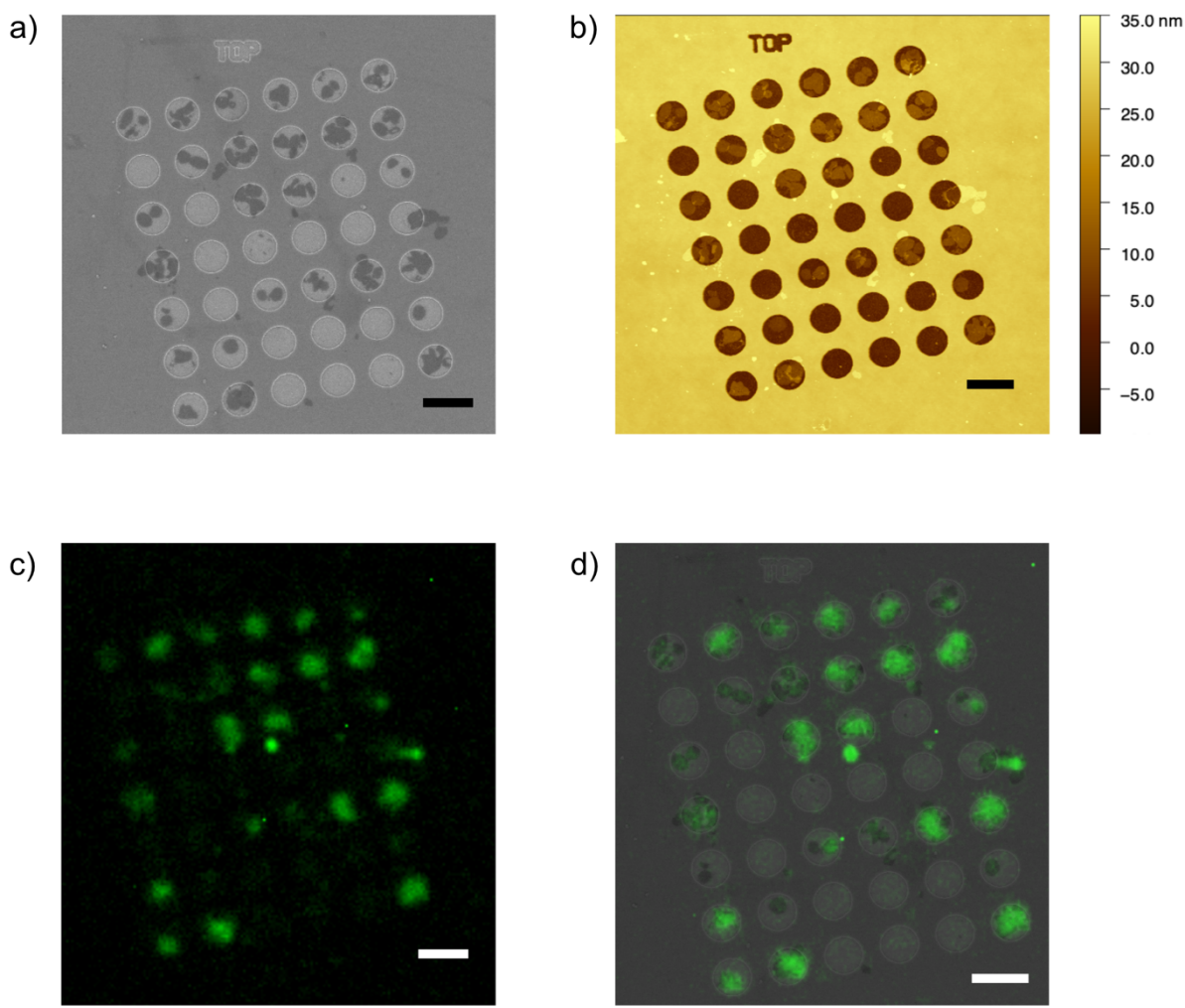

**Figure 3. Observation of PM patches after deposition in Type B (alumina bottom) traps.** (a) Scanning electron microscopy (SEM) image of an array of circular traps, many of them containing deposited PMs with C-terminal His-tag (C-His). Membranes exhibit a darker contrast. (b) AFM image of the same array as in a). (c) Fluorescence microscopy image of the same array. Membranes with anti-His-antibody show green fluorescence. (d) Overlay of fluorescence and SEM image demonstrating that most of the deposited membranes are presenting the His-tag and being recognized by the antibody. Scale bars are 2 μm.

**Traps formed entirely of SiO₂ (Type A).** In the traps entirely made of $SiO_2$ (Type A), the membranes were consistently deposited with the His-tag facing the bottom of the trap, *i.e.,* displaying no fluorescence. For the analysis, we only counted traps with either a single membrane patch or several membrane patches that were not overlapping or folded over. Traps containing overlapping membrane patches or aggregates of membranes were easily identified in the AFM scans through the height of the membrane structures. These frequently displayed fluorescence, because part of the membrane (or the PM patch on top of a first membrane) was positioned with the His-tag facing upward. Sometimes, aggregates of several membrane patches were trapped and displayed fluorescence because of the high probability of some His-tag surface of the aggregate being accessible to the antibodies. Such traps were also excluded from the evaluation. However, such multilayer or aggregate assemblies, typically fluorescent, were a useful internal control of the functionality of the antibody assay. Figure 4 shows the ratios of fluorescent and dark single layer PM patches for the different traps and membrane types. In the $SiO_2$-bottomed traps, only 1% of membrane patches were found to be fluorescent for both membrane variants (C-His or N-His).

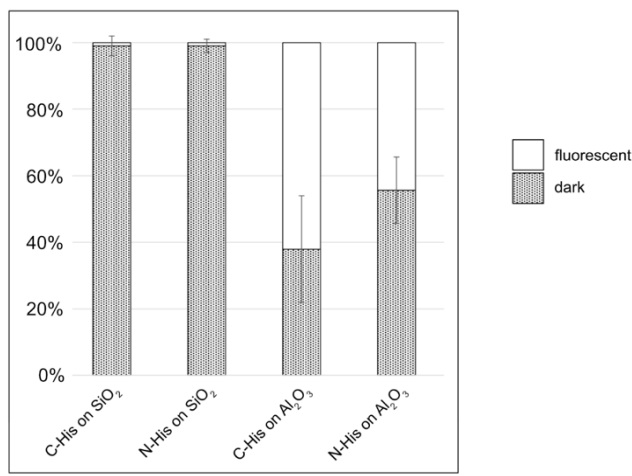

**Figure 4. Distribution of PM orientation in different trap types.** Graph showing the ratio of dark and fluorescent membranes in the array traps, indicating their orientation (fluorescent = His-tag up; dark = His-tag down). Each bar represents at least 100 membranes from at least seven trap arrays and from at least 2 independent deposition experiments. Error bars show the standard deviation of the averaged results of all trap arrays for each combination of trap type and membrane type. The comparison clearly shows that silica surfaces induce deposition with the His-tag facing down in almost every case, while the alumina surface promotes a much larger fraction of His-tag facing up.

**Traps with an $Al_2O_3$ bottom (Type B).** Substrates for the preparation of alumina-bottom trap arrays comprise a sandwich layer structure with 15 nm of alumina and 20 nm of $SiO_2$ on top. t-SPL-patterned trap arrays were transferred by RIE into the $SiO_2$ with the alumina serving as a stop layer, thus exposing alumina exclusively at the bottom of the traps (for details see Methods).
Alumina surfaces commonly have their isoelectric point at higher pH values (pH 7 - 9)[37] than $SiO_2$ and are thus expected to carry a positive surface charge in water at pH values < 7. We expected the alumina bottom traps to be more attractive for membrane surfaces of a larger negative potential and therefore less attractive for the His-tag labelled side of the membranes (scheme Figure 2b). Nevertheless, traps with $Al_2O_3$ on the bottom did not cause any direct adsorption of membranes from solution by opposite-charge attraction. This was substantiated by the lack of membranes adsorbed in the trapping sites of control fields, which were not directly under the pillar ($2h \gg 1$ μm, and $\psi_m \sim 0$) during the deposition process. This indicates that the alumina surface does not have a strong positive surface potential under our experimental conditions. We performed zeta-potential measurements of alumina surfaces prepared by atomic layer deposition (ALD) at pH levels similar to those of the trapping experiments and found values between 20 and 35 mV. However, surface zeta potential may vary depending on the exact process of preparation and might be less positive in the traps after the RIE process that was used to remove the $SiO_2$. Still, we consistently observed a stronger confinement of the membrane patches in $Al_2O_3$-bottom traps than in $SiO_2$ traps at comparable slit gaps (data not shown), suggesting a higher trapping potential of the $Al_2O_3$-bottom traps.

Permanent deposition of the membranes in the alumina-bottom traps was achieved by completely closing the gap. Comparing the ratio of fluorescent and dark membranes (Figure 4) deposited in alumina-bottom traps to those of the silica surface traps, one can see a significant increase of fluorescent PMs, *i.e.,* membranes exposing their His-tag on the top side. Note that the measured ratios of fluorescent PMs are only a lower boundary estimate for the His-tag-up orientation because the antibody assay has a limited efficiency. A small number of membrane patches (< 10%) might not carry His-tags and the binding efficiency of the antibody to the His-tag is not 100%. To estimate the efficiency of the antibody assay, we performed an efficiency test in bulk solution (for details see Methods). We found efficiency values of 63% and 70% for the antibody binding to the N-terminal and the C-terminal labelled membranes, respectively. Thus, the real values for alumina-bottom traps, estimated by the conditional probability rule, could be as high as 70% and 89% for the N-His and the C-His bR PM versions, respectively. Of course, this also must be considered for the ratios of orientation in silica surface traps, but even when accounting for the limited efficiency of the antibody assay in the same way, the fraction of His-tag-up membranes does not increase above 2%.

**Simulations.** It is remarkable that the position of the His-tag (N- or C-terminus) is the critical factor for the final orientation of the membranes and that the charge asymmetry that is already present in the wild type[10] is overridden by the His-tag. This provides us with a handle to achieve both membrane orientations, *i.e.,* pumping direction of the bR, by choosing the necessary combination of His-tag modification and surface potential. However, the symmetric charge conditions of the walls for the $SiO_2$ traps (see Figure 2a) should not lead to a preferred orientation of the membrane. In order to understand the observed asymmetric deposition in more detail, we employed finite element simulations using Fenics.[38]

We modeled the membrane as a spherical disc of 200 nm radius and 6 nm thickness in a weak electrolyte with a Debye length of 20 nm, resembling experimental conditions as determined by conductivity measurements. The surface charge of the membrane was set to be -2 electrons per protein area (11.4 nm$^2$) on one side[10] and +0.1 electrons per protein area on the His-tagged side. For the charge of the confining surfaces we assumed symmetric charging of the glass and the silica surface of 0.5 mC m$^{-2}$, consistent with the results of Behrens et al.[22] More details can be found in the Supporting Information. Figure 5 shows the result of the simulations.

For the symmetric charge boundary conditions of the $SiO_2$ traps, we indeed find similar trapping energies of 13.8 $k_BT$ (cyan line) and 13.3 $k_BT$ (blue line) for the membrane with the negative charge facing up or down, respectively. The magnitude of the trapping energies is consistent with the experimental observations that the membranes are stably trapped, but their difference of only 0.5 $k_BT$ cannot explain the observed orientation control of 99%. Note, however, that the equilibrium height of the membrane above the well is determined by its orientation. The positively charged side weakens the repulsion of the overall negatively charged membrane from the wall and causes the membranes to reside close to the wall facing this interface. Upon further reduction of the gap distance the free energy minimum disappears and is replaced by a gradient towards the closer wall causing the membranes to deposit there. Thus, the simulation suggests that all membranes with a high (negative) charge facing upward are deposited on the substrate (*i.e.,* in the trap) while membranes with the opposite orientation are deposited onto the glass pillar, consistent with the observed deposition asymmetry. Indeed, we found PMs deposited on the pillar when examining the pillar surface by AFM after the deposition experiment (data not shown), corroborating this interpretation.

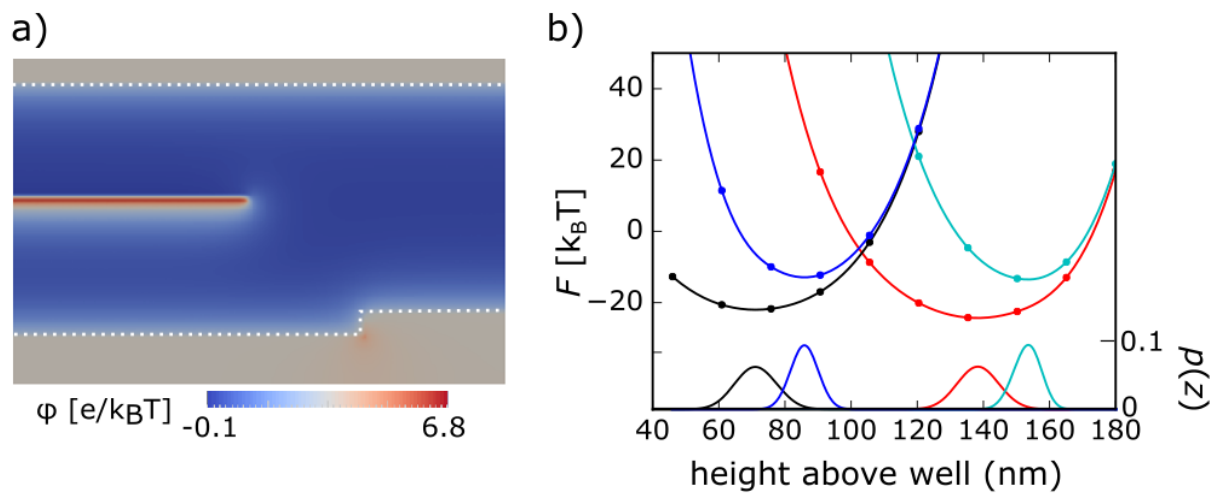

**Figure 5. Finite element simulation results.** a) The electrostatic potential of a membrane trapped in a well of 300 nm radius and 20 nm depth. The gap distance is 200 nm. The white dotted lines indicate the position of the walls. b) Free energy curves of the membrane positioned over a silica well (blue and cyan) and alumina well (black and red). The negatively charged side of the membrane is facing the well for the cyan and the red curves. The curves at the bottom depict the simulated probability density of the vertical position in the gap.

In the case of alumina at the bottom of the trap the situation is different. We find in the simulations that positive surface charges on the substrate always provide an unstable situation without minima in the free energy curves. Therefore, we assumed a weakly negative alumina charge density of -0.003 mC m$^{-2}$. In this case the trapping energies are -25.0 (red curve) and - 22.8 $k_BT$ (black curve), providing an energy gain of 2.2 $k_BT$ for the experimentally favored orientation of the negative charge facing down (red curve). The other orientation is still modeled to reside very close to the substrate and it would be transferred onto the substrate upon decreasing the gap distance. The observed results can be explained, if in this case all membranes are deposited on the substrate due to the weak charge on the substrate. Moreover, the energy difference of 2.2 $k_BT$ is larger than the thermal energy and causes flipping of a fraction of wrongly oriented membranes. This is consistent with the experimentally observed weaker control of the orientation in the case of alumina traps. Moreover, the number of deposited membranes per deposition step was higher on alumina compared to silica, which corroborates the interpretation that all trapped membranes were deposited on the substrate.

**Conclusion**

We have shown that nanofluidic confinement in combination with an ISCAT setup can be used for the label-free imaging and manipulation of biological entities. Our objects of interest, PMs, are characterized by a directional functionality (light-driven proton pump), hence control of their orientation is crucial for their activity. We were able to trap, transport and immobilize PM patches at pre-defined trapping sites. Tuning of the materials used to fabricate the lithographic structures, employed for electrostatic confinement, yielded unprecedented fine control over the membrane orientation within the trapping site.

The nanofluidic apparatus, combined with the tunability of both geometry and surface properties of the trapping substrate, constitutes a powerful tool for the manipulation and integration of supramolecular bio-structures into top-down-fabricated patterns, an essential prerequisite for the development of functional hybrid devices.

**Methods**

All chemicals and solvents without a special supplier specification were purchased from Sigma-Aldrich. The pH was measured with a handheld pH Meter (Horiba). Zeta-potential measurements were performed with a Zetasizer Nano ZS (Malvern Panalytical).

**Substrate fabrication.** The samples in our experiments originated from a highly doped silicon wafer and were diced to a size of 8 mm x 8 mm. We fabricated two types of trapping devices, Type A and B. Type A were additionally thermally oxidized to a thickness of 500 nm and featured a silica surface throughout the array of trapping sites. We would like to point out that silica is expected to develop a negative charge when in contact with an aqueous solution at neutral pH. In the next step, Type A samples were coated with a solution of 9 wt% of the thermally sensitive resist polyphthalaldehyde (PPA) (Phoenix 81, Allresist), dissolved in anisole. The PPA solution was spin coated at 3000 rpm and annealed at 110°C for 2 min, resulting in an 80 nm thick layer of PPA. We defined the trap locations via thermal scanning probe lithography (t-SPL), using a home-built instrument[29] or the commercial instrument Nano Frazor Scholar (Heidelberg Instruments Nano).

The PPA was patterned via a heated tip (tip heater nominally between 950 and 1250°C), capacitively pulled into contact with the substrate. The local decomposition of the PPA resulted in cylindrically shaped indentations, all the way through the polymer. Each device contained arrays of 9x9 or 7x6 cavities, 1.2 - 1.5 µm in diameter, arranged with a spacing of 2 µm. After an oxygen descum, the pattern was transferred to a depth of 20 nm in the silica layer, using Reactive Ion Etching (RIE). We utilized a mixture of $CHF_3$, argon and oxygen gases (12, 38 and 1 sscm, respectively) with a power of 100 W and 30 mbar chamber pressure.

Type B devices were fabricated by covering the bare silicon wafer with alumina using atomic layer deposition (ALD, Picosun). We deposited 15 nm of alumina at 300°C, and immediately covered this layer with 20 nm of $SiO_2$, also via ALD (Oxford Instruments). After this step, the devices were covered by PPA and patterned as described above. The RIE etch in this case ensued the complete removal of silica from the bottom of the voids, and resulted in exposed alumina, which is expected to attain a positive charge in contact with an electrolyte at pH 7. The thickness of silica essentially defined the depth of the indentations since the alumina film acts as a stopping layer when using the same RIE recipe as for Type A devices.

Type A and B devices can be re-used after a cleaning procedure: The devices were sonicated for 20 min in 1% sodium dodecyl sulfate (SDS) solution. After washing with copious amounts of DI water they were dried and treated with a 200 W oxygen plasma twice for 30 s with a break of 1 min (Tepla AG).

**Engineering, homologous expression and isolation of PMs containing N- and C-terminally $His_{10}$-tagged bacteriorhodopsin.** A C-terminal $His_{10}$-tag was introduced into the gene encoding bacterio-opsin (*bop*) by polymerase chain reaction (PCR) using the forward primer 5'-AAA AGG ATC CGA CGT GAA GAT GGG GC-3' and the reverse primer 5'-AAA AAA GCT TGA TTC AGT GGT GAT GAT GGT GAT GAT GGT GGT GAT GTC CGT CGC TGG TCG CGG CCG CGC-3'.[35] The N-terminal $His_{10}$-tag was engineered as described previously.[34] The genetically engineered bR variants were cloned into the shuttle vector pHS blue and transformed and expressed homologously in the *bop*-deficient *H. salinarum* strain L33.[35,39] Isolation of PMs containing the engineered BR variants was performed using established protocols.[9,35,39–41] Briefly, *H. salinarum* cells were harvested and lysed by osmotic shock in ultrapure water. After removal of cell debris by three centrifugation steps (1x 4300*g*, 10 min, 4°C; 2x 7600*g*, 10 min, 4°C), PMs were isolated using four rounds of ultracentrifugation (2x 55000*g*, 1 h, 4°C; 2x 60000*g*, 1 h, 4°C) followed by resuspension in ultrapure water. The concentration of bR in the sample was determined by chromophore-based absorption spectrometry at 568 nm using a molar extinction coefficient of 63,000 l $mol^{-1}$ $cm^{-1}$.[42] The His-tagged PMs are very stable for several months. Periodic routine quality analyses of His-tagged PMs by sodium dodecylsulfate-polyacrylamide gel electrophoresis (SDS-PAGE) analysis indicate no degradation of bacteriorhodopsin over time and by exposure to day light.

Membranes were dispersed in ultrapure water at a concentration of 0.1 – 0.2 mg $ml^{-1}$ without any further addition of salts or buffer. The dispersion was stable over weeks in the refrigerator. Before use, pre-diluted aliquots were taken from the refrigerator, allowed to reach room temperature and vortexed for a few seconds to disperse settled membranes.

**Antibody assay.** We removed the sample from the nanofluidic confinement stage, carefully removing excess liquid without completely drying the substrate. In order to avoid unspecific deposition of the antibody on the substrate surface, we then applied SuperBlock (Thermo Fisher) solution. The SuperBlock solution was left on the sample for 10 min. Then the liquid was carefully removed again without complete drying and antibody solution (20 µl, His-tag monoclonal antibody-Alexa Fluor 488, Thermo Fisher) (20 μg ml$^{-1}$ in PBS, pH 7.4) was applied to the pattern for 30 min. Next, the sample was rinsed with ultrapure water (2 x 100 μl) and dried in a gentle stream of argon. Sample arrays were first analyzed with the fluorescence microscope (Axiotech, Zeiss), followed by the capture of AFM images (Dimension V, Bruker) for the same fields. In one case scanning electron microscopy (SEM) images (Leo 1550, Zeiss) were taken as well. Fluorescence images were captured with μManager[43] and analyzed with ImageJ.[44] AFM images were edited and analyzed with Gwyddion.[45]

**Antibody efficiency test.** We incubated one sample each of the C-terminal His-tag and the N-terminal His-tag PMs with the antibody solution. For this, we mixed the corresponding membrane stock solution (25 μl, 0.2 mg ml$^{-1}$) with the antibody stock solution (50 μl, 0.02 mg ml$^{-1}$). After 1 h, any excess or unbound antibody was removed by 4 cycles of centrifugation (4000 rpm, 20 min, Eppendorf 5424) and re-dispersion in ultrapure water. Next, a drop of the solution was left to dry on an alumina surface. After drying, the samples were analyzed by fluorescence microscopy and SEM. An overlay of the two imaging techniques provides a ratio of fluorescent to non-fluorescent membranes. In this control experiment all His-tagged surfaces are accessible to the antibodies, because the binding takes place in solution. Also, the fluorescence of the membranes does not depend on their orientation, because the binding of the antibody takes place before the adsorption to the surface.

**Statistical Analysis.** For each combination of trap type (SiO$_2$, Al$_2$O$_3$) and membrane type (C-His, N-His) at least two independent experiments, together representing at least 7 trap arrays and at least 100 membranes were evaluated for the ratio of dark and bright membranes. The ratios of the individual trap arrays for one combination of trap type and membrane type were averaged and the result, with standard deviation as error bars, plotted in Figure 4.

**Filter to enhance contrast in the ISCAT images.** Due to the small difference in refractive index of the membranes and the surrounding fluid, the membranes had low contrast in the ISCAT images. To enhance their visibility, we first recorded a 20 s long video of trapped membranes, probing the trap potentials at a frame rate of 100 fps. In the next step, we determined the temporal median of the video which eliminated all moving objects, *i.e.,* the membranes, and delivered a static background. Then, we took a single frame from the middle of the video and divided it by the static background. The resulting image was then rescaled such that it encompassed a full range of 12bit again. In the last step, the rescaled image was overlaid with the static background at a ratio of 30 to 70.

**FEM modelling.** We performed all computations on a desktop computer using finite element software. The calculation was performed using the Fenics[38] software package. Meshes were prepared using Gmsh[46] and solutions were inspected using Paraview.[47] For more details, see SI.

**Data Availability**

Data available from the authors upon reasonable request.

**Acknowledgements**

We thank the Cleanroom Operations Team of the Binnig and Rohrer Nanotechnology Center (BRNC) for their help and support. Funding was provided by the European Research Council (StGno. 307079, PoC Grant no. 825794) and by the Swiss National Science Foundation (SNSF; grant 200021_179148). This work is part of a project that has received funding from the European Union's Horizon 2020 research and innovation programme under the Marie Skłodowska-Curie grant agreement No 812868. D.F. acknowledges funding from the University of Bern, the Swiss National Science Foundation (SNSF; grant 310030_184980) and the Swiss National Centre of Competence in Research (NCCR) Molecular Systems Engineering.

**Author contributions**

M.S., P.D.B., and D.F. prepared PM patches. F.R., C.S., P.M.N., J.F.S., K.K., A.K., and H.W. performed experimental work in trap fabrication, nanofluidic confinement and analysis of membrane orientation. H.W., C.S., and A.K. analyzed acquired data. A.K. performed FE simulations. All authors contributed to manuscript preparation.

**Competing interests**

The authors declare no competing interests.